\documentclass[twocolumn]{aastex61}

\newcommand{\ktwo}{\textit{K2}\,}
\newcommand{\vsini}{$v\sin i$\,}

\accepted{for publication in ApJ Letters}

\shorttitle{K2 reveals pulsed accretion driven by CI~Tau b}
\shortauthors{Biddle et al.}

\begin{document}

\title{K2 reveals pulsed accretion driven by the 2 Myr old hot Jupiter CI~Tau b}

\correspondingauthor{Lauren I. Biddle}
\email{lbiddle@lowell.edu}
\author[0000-0003-2646-3727]{Lauren I. Biddle}
\affiliation{Lowell Observatory, 1400 W. Mars Hill Rd. Flagstaff. Arizona. 86001. USA.}
\affiliation{Department of Physics \& Astronomy, Northern Arizona University, 527 S. Beaver St. Flagstaff. Arizona. 86011. USA.}

\author{Christopher M. Johns-Krull}
\affiliation{Department of Physics \& Astronomy, Rice University, 6100 Main St. MS-108, Houston, TX 77005}
\author[0000-0003-4450-0368]{Joe Llama}
\affiliation{Lowell Observatory, 1400 W. Mars Hill Rd. Flagstaff. Arizona. 86001. USA.}
\author{Lisa Prato}
\affiliation{Lowell Observatory, 1400 W. Mars Hill Rd. Flagstaff. Arizona. 86001. USA.}
\author{Brian A. Skiff}
\affiliation{Lowell Observatory, 1400 W. Mars Hill Rd. Flagstaff. Arizona. 86001. USA.}




\begin{abstract}
CI~Tau is a young ($\sim$2 Myr) classical T Tauri star located in the Taurus star forming region. Radial velocity observations indicate it hosts a Jupiter-sized planet with an orbital period of approximately 9 days. In this work, we analyze time series of CI~Tau's photometric variability as seen by \ktwo. The lightcurve reveals the stellar rotation period to be $\sim$6.6 d. Although there is no evidence that CI~Tau b transits the host star, a $\sim$9 d signature is also present in the lightcurve. We believe this is most likely caused by planet-disk interactions which perturb the accretion flow onto the star, resulting in a periodic modulation of the brightness with the $\sim9$ d period of the planet's orbit.
\end{abstract}




\section{Introduction} \label{sec:intro}
Detecting and characterizing exoplanets around very young stars enables a direct view into the formation and evolution of planetary systems. Mature planetary systems encompass a range of configurations including hot Jupiters orbiting within tenths of an AU from their host stars and super-Earths orbiting within the habitable zone of low mass stars \citep{Howard2010,Borucki2011,Kopparapu2013,Burke2014,Winn2015}. While there are now nearly 3000 confirmed exoplanets\footnote{\url{http://www.exoplanets.org} (2018 Jan. 12)} around main-sequence stars, post-main sequence stars, brown dwarfs, and pulsars, only a handful of candidates have been found around the youngest of stars. One reason for the lack of detected planets around young stars is simply that there are very few nearby star forming regions, with most being beyond 120 pc. Consequently, radial velocity (RV) measurements become difficult because of the faintness of the host stars.  Even for bright young stars, exoplanet detection remains difficult in relatively nearby star-forming regions such as Ophiuchus and Taurus. Young classical T Tauri stars typically have optically thick, actively accreting circumstellar disks and exhibit stellar activity signatures caused by cold star spots as well as variable accretion of disk material channeled onto the star along strong magnetic field lines \citep{herbst2007}. These astrophysical phenomena manifest themselves in both photometric and RV variability studies, making exoplanet detection around young stars extremely challenging \citep{Queloz2001,Bouvier2007,Robertson2014}.

Despite these challenges, there have been a number of recently announced exoplanets around T Tauri stars, detected through a variety of methods, including V830 Tau b \citep{donati2016}, K2-33 b \citep{mann2016,david2016}, and TAP 26 b \citep{yu2017}. Our team reported the RV detection of CI~Tau b  \citep{johnskrull2016}, a 2 Myr old hot Jupiter around a classical T Tauri star. CI~Tau b was detected after an extensive radial velocity campaign combining both optical and infrared measurements in an effort to disentangle the planet signal from stellar induced RV signals. From the RV measurements, \citet{johnskrull2016} determined the minimum mass of the planet to be $M_p\sin i=8.08\pm1.53 M_{\rm Jup}$ and the orbital period to be $P_{\rm orb} = 8.9891\pm 0.0202$ d.  Under the assumption that the stellar inclination is equal to that of the circumstellar disk ($i$ = $45.7\pm1.1^{\circ}$, \citealt{guilloteau2014}), then $M_p = 11.29\pm2.16 M_{\rm Jup}$.  In addition, \citet{johnskrull2016} acquired ground-based photometry to probe CI~Tau's photometric variability, which yielded a stellar rotation period of 7.1 d, though with considerable uncertainty.

A new opportunity to characterize CI~Tau's photometric variability was provided by \ktwo, the extended mission of the \textit{Kepler} space telescope \citep{borucki2010}, intended to observe various fields in the ecliptic plane for periods of approximately 80 days \citep{howell2014}. In this letter, we present the \ktwo photometry of the CI~Tau system released on August 28, 2017, which provides further evidence for the presence of CI~Tau b, enabling us to disentangle the stellar rotation signal from that of the planet-disk interaction.

\section{Observations and Analysis}

\begin{figure*}[t]
\centering
    \includegraphics[width=1\textwidth]{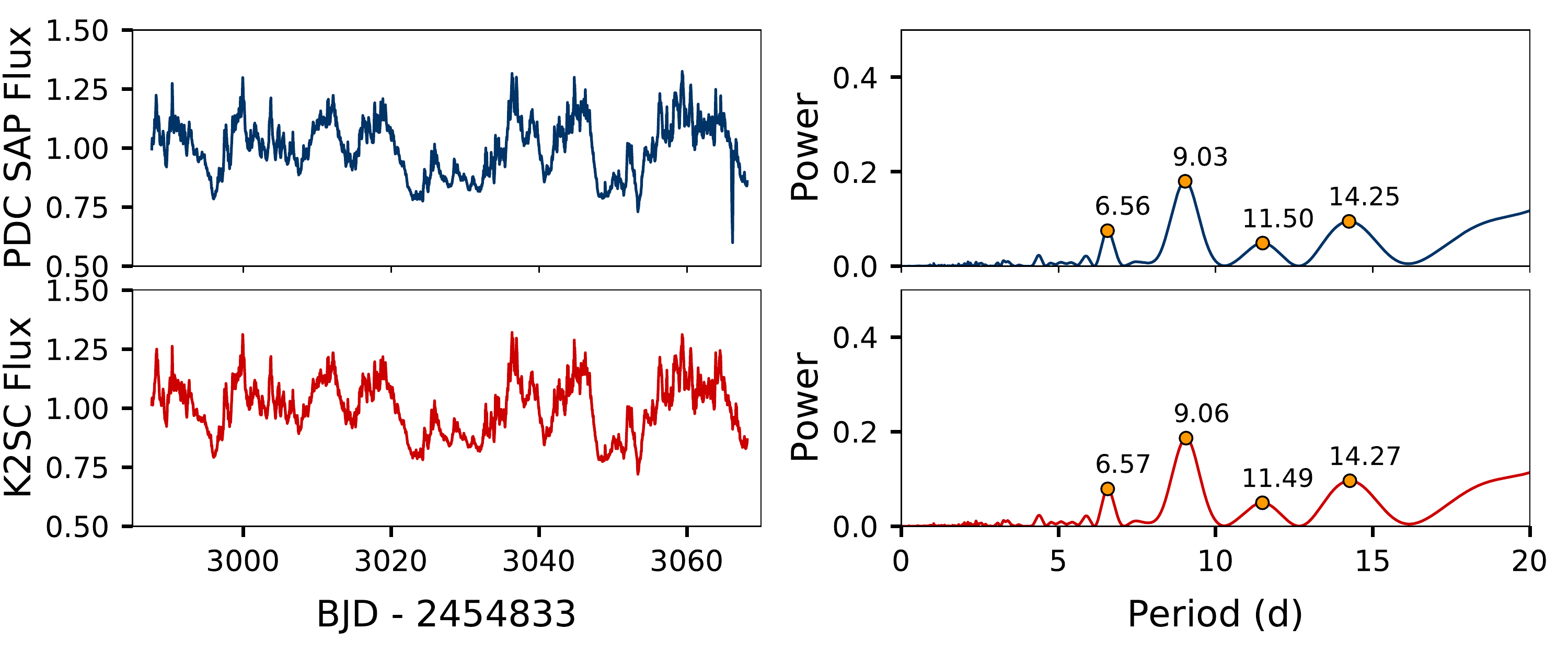}
    \caption{Lightcurves and periodograms of CI~Tau. The top row displays the PDC\_SAP lightcurve on the left and its corresponding periodogram on the right.  Peak signals in the periodogram are marked with gold circles and their respective periods. The bottom row displays the same for the K2SC lightcurve.}
\label{fig:lc}
\end{figure*}

\ktwo long-cadence time-series photometry of CI~Tau (EPIC 247584113) was acquired during Campaign 13 between 8 March 2017 and 27 May 27 2017 UTC. We downloaded the lightcurve from the Milkulski Archive for Space Telescopes (MAST\footnote{\url{https://archive.stsci.edu/k2}}). We detrended the PDC\_SAP \citep{jenkins2010} lightcurve with the K2SC pipeline \citep{aigrain2015,aigrain2016}, which uses Gaussian Process regression to remove instrumental systematics in the data while preserving astrophysical variability of the host star.  The top left-hand panel of Figure \ref{fig:lc} shows the PDC\_SAP flux provided by the \ktwo team and the post-processing result from the K2SC pipeline is shown in the bottom panel.  Each lightcurve exhibits approximately $\sim40\%$ variability on multiple time scales within the 80 days of observations. The source of this variability is from surface inhomogeneities likely resulting from the combination of cool spots and accretion hot processes.

For each lightcurve, we computed a generalized Lomb-Scargle periodogram \citep{zechmeister2009} using the \textsc{python} AstroML\footnote{\url{http://www.astroml.org/index.html}} package. Each periodogram was computed between 0.042 and 80 days (consistent with the Nyquist sampling frequency; \citealt{press1992}). We searched 10,000 points within this window to ensure that the peaks in the periodogram were well resolved. These limits and samplings are consistent with the recommendations of \citet{vanderplas2017}. The periodograms show distinct signals at 6.56$\pm$0.22 d and 9.03$\pm$0.51 d for the PDC\_SAP flux and at 6.57$\pm$0.24 d and 9.06$\pm$0.50 d for the K2SC flux (Figure \ref{fig:lc}, \textit{right}).  No strong peaks are seen at periods beyond 30 d; these are not shown.  The period uncertainty was estimated from the full width at half of the maximum of the periodogram power distribution surrounding each respective period \citep{ivezic2014}. We calculated the false alarm probability (FAP) using both the analytic solution \citep{zechmeister2009} and a Monte Carlo bootstrap algorithm for both periods in each lightcurve. Both methods yield a FAP of $<10^{-6}$ for both periods. For completeness, we conducted identical analyses of detrended lightcurves from the EVEREST \citep{luger2016,luger2017} and K2SFF pipelines \citep{vanderburg2015}, and the results from each were consistent with both the PDC\_SAP and K2SC.

\begin{figure*}
\centering
    \includegraphics[width=0.80\textwidth]{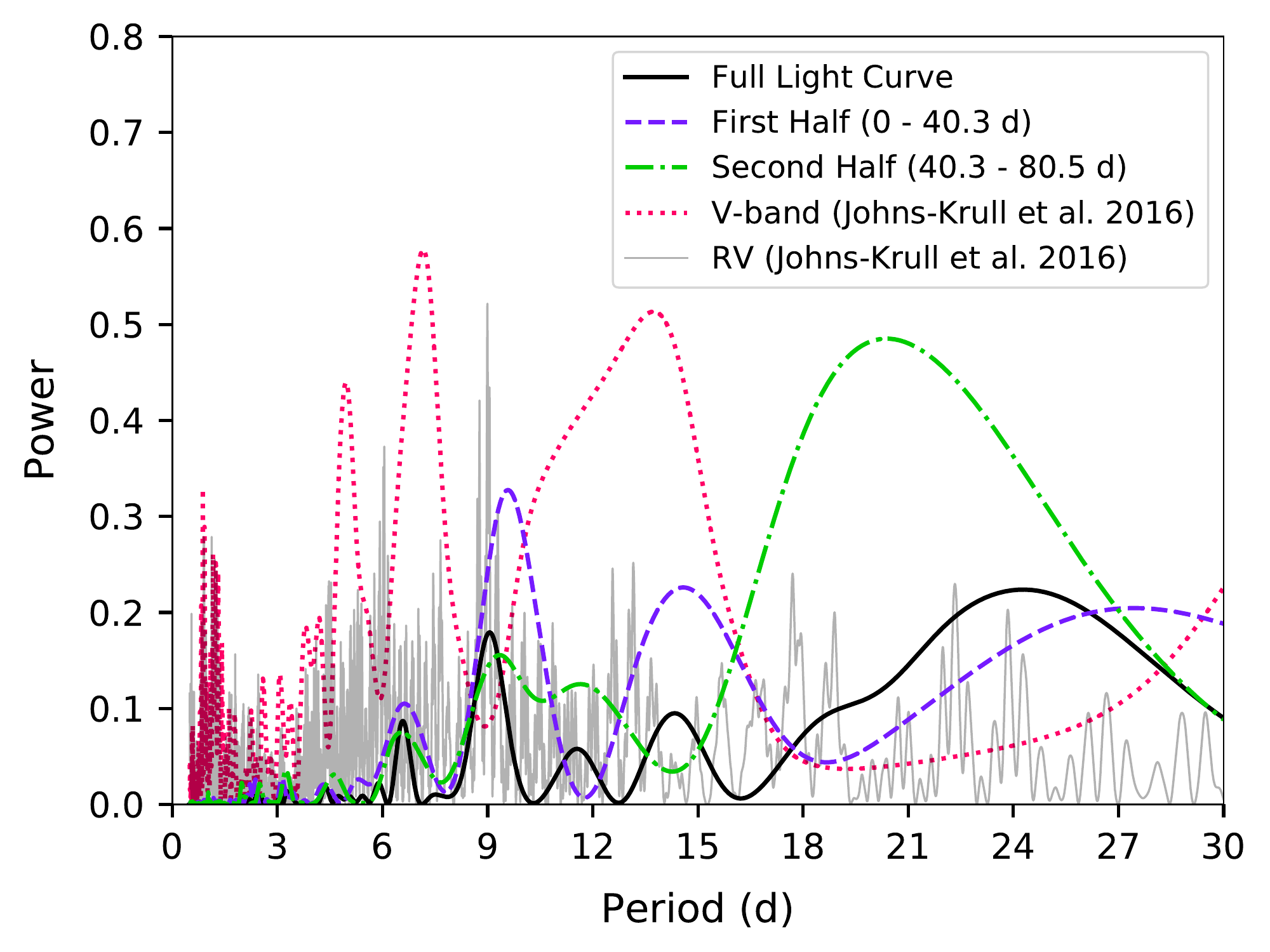}
    \caption{A direct comparison of the individual periodogram analyses of the full, first, and second halves of the K2SC lightcurve to an identical re-analysis of both the ground-based V-band photometry and combined visible and infrared RVs presented by \citet{johnskrull2016}. The first half shows periodicity at 6.73$\pm$0.60 d and 9.50$\pm$0.97  d, while the second half shows at 6.62$\pm$0.51 d and 9.41$\pm$0.95 d. The PDC\_SAP lightcurve yielded similar results (6.81$\pm$0.60 d and 9.47$\pm$0.96 d for the first half and 6.50$\pm$0.47 d and 9.41$\pm$0.96 d for the second).}
\label{fig:halfcompare}
\end{figure*}

To check whether the two periods hold throughout the series, we performed periodogram analyses on the first and second halves of the data sets separately (Figure \ref{fig:halfcompare}).  The $\sim$6.6 d and $\sim$9.0 d signals were present in both results.  The locations of the $\sim$6.6 d periods as well as their strength remained relatively constant compared to the $\sim$9.0 d signal, which is clearly stronger during the first half of the observing period and exhibits slightly larger values ($\sim$9.5 d) compared to the full lightcurve ($\sim$9.0 d).  We also calculated the FAP for the $\sim$6.6 d and $\sim$9.0 d periods for the first and second half lightcurves, results of which also returned $<10^{-6}$. The increased width of the peaks in the power spectrum at $\sim$6.6 d and $\sim$9.0 d when looking at the first or second half of the data is attributable to the lower frequency resolution that results from the decrease in the length of the time series when looking at only half of the data.

\begin{figure*}
\centering
    \includegraphics[width=1\textwidth]{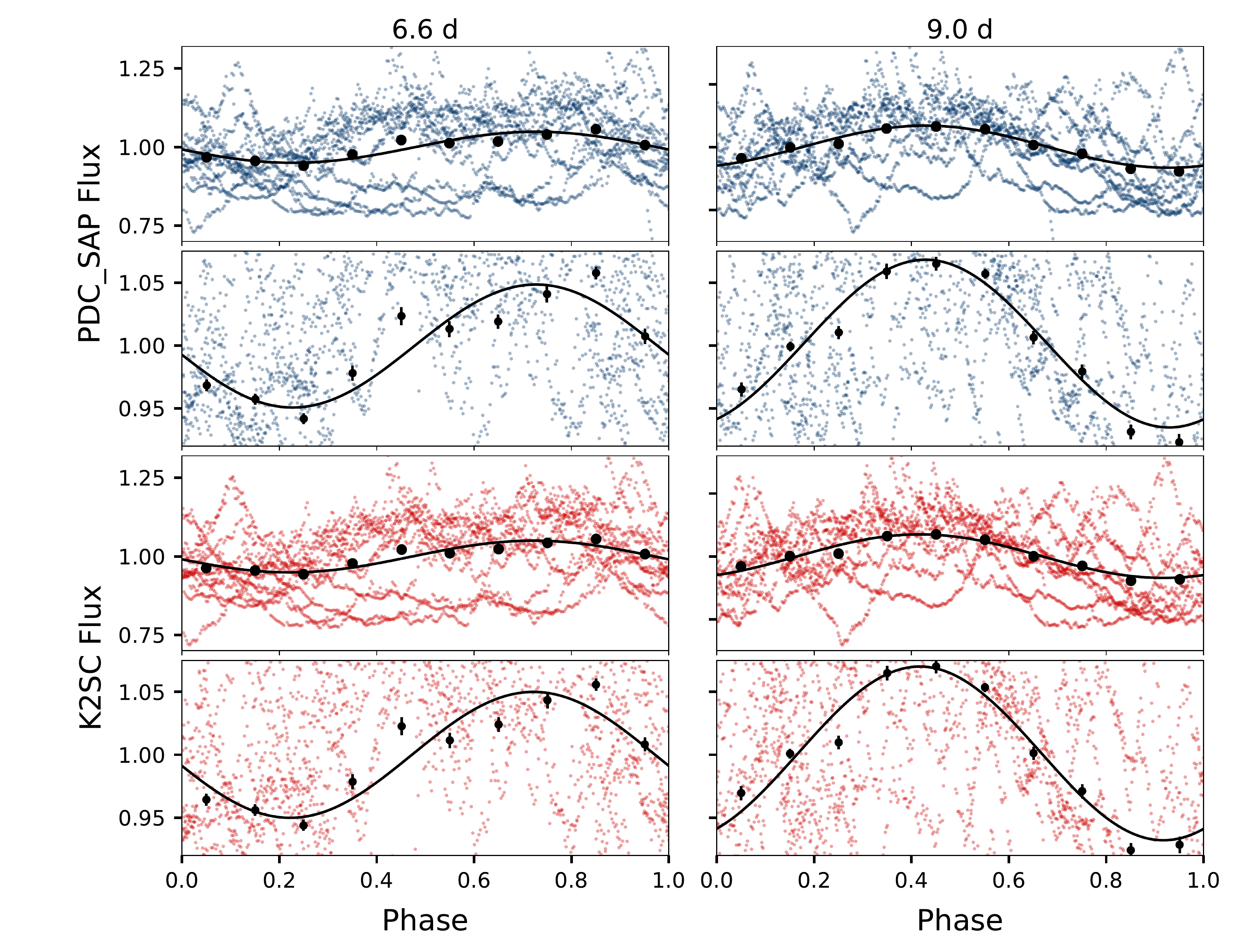}
    \caption{The left column shows the PDC\_SAP lightcurve (top two rows) and K2SC lightcurve (bottom two rows) phased to their respective $\sim$6.6 d periods. The column on the right shows the lightcurves phased to their respective $\sim$9.0 d periods. The black circles represent the data binned to 0.1 in phase. The sine fits to the binned data are also displayed in black. The second and fourth rows scale the y-axis to the amplitude of the binned flux to aid the eye.}
\label{fig:phase}
\end{figure*}

We constructed lightcurves phased to the $\sim$6.6 d and $\sim$9.0 d periods (Figure \ref{fig:phase}). The phased lightcurves appear to be coherent with both periodic signals. While there is considerable scatter in the phased plots, the points are concentrated in a way that appears periodic, which can be seen in the first and third rows of Figure \ref{fig:phase}. To better show this signal, we binned the photometric data points to 0.1 in phase, computed the mean and uncertainty of the mean within each bin, and then performed a sine fit to the binned data to measure the amplitude and significance of each signal. The fit to the $\sim$6.6 d phased lightcurves yielded an amplitude of 0.049$\pm$0.008 normalized flux units for the PDC\_SAP lightcurve (a $\sim$6 sigma signal) and 0.050$\pm$0.007 normalized flux units for the K2SC lightcurve (a $\sim$7 sigma signal). The lightcurve sine wave amplitude of the $\sim$9.0 d fit for each lightcurve was 0.067$\pm$0.006 ($\sim$11 sigma) and 0.069$\pm$0.006 ($\sim$11.5 sigma) in normalized flux, respectively. The amplitudes are large compared to their uncertainties, which further reinforces the significance of both signals.

In Figure \ref{fig:lc}, other peaks are apparent at 11.5 d and 14.3 d; however, these peaks are not consistently evident when the data are examined in the two subsets illustrated in Figure \ref{fig:halfcompare}.  When considering only half the data, the number of periods of these potential signals that is covered is reduced. For the 11.5 d peak, 3.5 periods are covered, and for the 14.3 d peak 2.8 periods are covered. However, for the 9 d peak, only 4.4 periods are covered: a 25\% increase compared to the 11.5 d peak and a 50\% increase on the 14.3 d peak. While this is relatively significant when considering the 14.3 d peak, it is fairly marginal for the 11.5 d peak, yet the 9 d peak survives strongly in all the periodograms of Figure \ref{fig:halfcompare}. There is also a significant peak in the 20$-$27 d period range in the analysis of the full light curve, and in the first and second half subsets.  Given the $\sim$80 d duration of the campaign, the full data set covers just over 3 full periods for the $\sim$24 d signal. It is thus unlikely that there is a true periodicity at 24 d. When we phase the full data set to a 24 d period, the result does not suggest a sinusoidal like that seen in Figure \ref{fig:phase}. Therefore, at this time we are not convinced that the 24 d period is real. We also note that for a period of 6.6 and 9.0 d, the light curve should show a beat period at exactly 24.75 days. Future compilation of all ground-based observations may shed light on the potential sources of these signals.

We obtained a new estimate of the $v$sin$i$ of CI~Tau using the optical echelle spectra presented in \citet{johnskrull2016}. In order to work with highest possible signal-to-noise data available to us, we used the least squares deconvolution (LSD) profiles that were computed by \citet{johnskrull2016} giving us a total of 26 observations of CI Tau with which to measure the $v$sin$i$. LSD \citep{donati1997} is a technique similar to cross correlation and is used to compute an average line profile. In the case of our CI~Tau spectra, a total of 1944 photospheric absorption lines were used to compute the average line which has a line-depth weighted mean wavelength of 6118 \AA. In order to estimate the $v$sin$i$ of CI~Tau we computed the LSD profile, using the same 1944 lines, of HD~65277 observed on 15 November 2013 with the same setup as the observations of CI~Tau \citep{johnskrull2016}. HD~65277 is a K5V star with a $v$sin$i \leq 0.53$ km s$^{-1}$ \citep{martinez2010}, and therefore represents an essentially non-rotating star of very similar spectral type to CI~Tau. To measure the $v$sin$i$ we then rotationally broadened the LSD profile of HD 65277, using a standard rotational broadening kernel (e.g., Gray 2008) with a limb darkening coefficient of 0.72, appropriate for the spectral type of CI~Tau (K7) and the wavelength (6118 \AA ) of the line being analyzed. We broadened the HD 65277 LSD profile with $v$sin$i$ values ranging from 1 to 20 km s$^{-1}$ in steps of 1 km s$^{-1}$ and measured the FWHM of the resulting broadened profile in order to define the relationship between $v$sin$i$ and FWHM. We then measured the FWHM of each of the 26 CI~Tau LSD profiles and interpolated on the defined relationship to determine the $v$sin$i$ of CI~Tau. Taking the mean and the standard deviation of the mean as the value of the random error, we find $v$sin$i = 10.08 \pm 0.14$ km s$^{-1}$ for CI~Tau. In order to estimate a systematic error we first repeated the process using limb darkening coefficients of 0.54 and 0.78, which represent the two extremes of the coefficients based on the full wavelength range of the lines used to compute the LSD profile.  Taking the largest difference with our original measurement, we assume this is the (likely over-) estimate (0.25 km s$^{-1}$) of the systematic error resulting from our choice of limb darkening coefficient and rotational broadening kernel. We then attempted to estimate the error that might result from using the LSD profiles themselves. We rotationally broadened the full observed HD~65277 spectrum by 10 km s$^{-1}$ and then computed the LSD profile of this broadened spectrum. We then measured the FWHM and used our same relationship to find its $v$sin$i$. Taking the difference (0.12 km s$^{-1}$) with the true value gives a measure of the systematic uncertainty caused by using the LSD method in the first place. We then add all three sources of uncertainty in quadrature to get a final measurement for the $v$sin$i$ of CI~Tau of $10.08 \pm 0.31$ km s$^{-1}$.

\section{Discussion}

From the Lomb-Scargle periodograms shown in Figures \ref{fig:lc} and \ref{fig:halfcompare}, it is clear that there are at least two significant astrophysical signals in the \ktwo lightcurve of CI~Tau at $\sim6.6$ d and $\sim9.0$ d. We identify the $\sim6.6$ d signal as the stellar rotation period. We use our \vsini measurement of CI~Tau, and the inclination of CI~Tau's disk, $i$ = $45.7\pm1.1^{\circ}$ \citep{guilloteau2014}, to estimate the stellar radius. Assuming the disk and stellar spin axes are aligned, we have
\begin{equation}
   {R_\star} = \frac{1.96\times10^{-2}\times P_{\rm rot}({\rm days})\times v\sin i({\rm km\,s}^{-1})}{\sin i}
\end{equation}
where $P_{\rm rot}$ is the period from the Lomb-Scargle periodogram analysis. For the peak at $P=6.6$ d this corresponds to a stellar radius, $R_\star=1.81\pm0.08R_\odot$, whereas the period at $P=9.0$ d would correspond to a stellar radius of $R_\star=2.50\pm0.15R_\odot$. From the luminosity of CI~Tau ($L_\star =0.81L_\odot$) and an effective temperature of 4,060 K \citep{mcclure2013}, we estimate the radius of the star from $L_\star=4\pi R_\star^2\sigma T^4_{\rm eff}$ to be $R_\star=1.81R_\odot$, consistent with the $\sim$6.6 d period but inconsistent with the $\sim$9.0 d period.  For a K7 star of age $\sim$2 Myr, a radius of $\sim$2.5 R$_{\odot}$ is unrealistic (e.g., \citealt{baraffe2015}).

With the identification of the $\sim$6.6 d stellar rotation period, the question remains, what is origin of the $\sim$9.0 d signal in the periodogram? Because of \ktwo's large pixel size, we considered the possibility of contamination from additional sources in the field and found that images from DSS, Galex, 2MASS, WISE, and PanSTARRS show that there are no objects of comparable brightness within an $1\arcmin$ of CI~Tau. We also examined the potential for multiplicity in the system. High resolution observations reaching 5$\sigma$ contrast at $0\farcs25$ separation \citep{uyama2017} provide no evidence for a companion down to $\Delta$H = 6.8 that could contribute to the photometric signal. However, there is evidence in support of a sub-stellar body. \citet{johnskrull2016} reported the detection of a planet orbiting CI~Tau using data from an extensive optical and infrared RV survey. The planet mass they derived is $M_p = 11.29\pm2.16 M_{\rm Jup}$ and the orbital period is $P_{\rm orb} = 8.9891\pm0.0202$ d, consistent with the $\sim$9.0 d period shown in both of the Lomb-Scargle periodograms in Figure \ref{fig:lc}. Our current understanding of the CI~Tau system is that the planet does not transit. This is supported by evidence that the disk is inclined $i\sim45^\circ$ \citep{guilloteau2014}, though it is possible that a planet could be in an orbit misaligned with disk mid-plane (e.g., Kepler-63; \citealt{sanchis2013}). However, the $\sim$9.0 d periodic signal may be the result of a planet-disk interaction because the presence of a massive planet in an actively accreting disk should show both spectroscopic and photometric variability. Indeed, \citet{johnskrull2016} find evidence in the H$\alpha$ profile variations of CI~Tau that the planet may be modulating the accretion of disk material onto the star. Although hot spots located at the foot of accretion streams can affect RV measurements mimicking the signal of an orbiting body \citep{kospal2014,sicilia2015}, \citet{johnskrull2016} specifically looked for these signals and found no evidence that hot spots produced the RV signals seen in photospheric absorption lines. Given the stellar mass of 0.80 M$_\odot$ \citep{guilloteau2014}, \citet{johnskrull2016} determined a semi-major axis of 0.079 AU for the planetary orbit. This would place the planet inside the inner edge of the disk at 0.12 AU \citep{mcclure2013}. It is probable that an 11.3 M$_{\rm Jup}$ planet so close to the inner edge of the disk would stimulate the accretion of material and possibly modify accretion onto the star, creating non-axisymmetric accretion flows \citep{tofflemire2017a,tofflemire2017b}. As the planet orbits the star, this interaction could produce a periodic variation in the H$\alpha$ emission, a tracer of accretion. The impact on stellar accretion could produce photometric variability on the $\sim$9.0 d period of the planet's orbit, which K2 data show as a periodic component in the system brightness. Additionally, the strength of this periodic component can be expected to fluctuate because of the sporadic nature of accretion on short timescales \citep{herbst2007}. This is observed in the analysis of the first and second halves of the data and is in contrast to the relatively consistent signal strength of the $\sim$6.6 d periodic component in brightness, which is caused by cold star spots that are long-lived in young stars (e.g., \citealt{stelzer2003}) and produce relatively consistent fluctuations in brightness as the star rotates \citep{herbst2007, bradshaw2014}. If the 9.0 d periodic signal is the result of stellar rotation, it would call into question the legitimacy of the planet. On the other hand, the similar RV amplitudes seen in the optical and the IR and the null results on tests for RV variations produced by an accretion hot spot \citep{johnskrull2016} attest to the significance of planet's detection. In addition, we would then need to explain the source of the 6.6 d signal which is the most persistent signal observed in Figures \ref{fig:lc} and \ref{fig:halfcompare}.

The exact mechanism by which a hot Jupiter might itself accrete and modulate disk accretion onto a young star has not been simulated to our knowledge.  However, there have been several simulations of accretion from circumbinary disks onto the central stars in the binary system \citep{artymowicz1996,gunther2002,munoz2016}.  Most of these simulations focus on equal or nearly equal mass stars in either circular or eccentric orbits. For equal mass stars in circular orbits, quasi periodic variations in the accretion rate are seen, but with a dominant period that is a few times the binary orbital period (e.g., \citealt{munoz2016}). For eccentric systems, the accretion is seen to vary with a period equal to the binary orbital period.  \citet{johnskrull2016} find a best fit solution to their RV measurements for an eccentric orbit of the planet around CI~Tau; however, the uncertainty in the eccentricity is large enough that the orbit could be circular. The mass ratio found by \citep{johnskrull2016} for the CI~Tau star and planet system is $\sim 0.014$. \citet{dorazio2013} simulate black hole binaries in circular orbits with a range of mass ratios from 1.0 down to 0.003.  At a mass ratio of 0.01 they find that visible accretion streams are still present with the stream corotating with the binary orbital period. Such a scenario would produce accretion modulated with a period equal to the binary orbit when viewed from a fixed location (e.g., the Earth). Furthermore, periodic accretion variability has been observed in eccentric T Tauri binaries \citep{tofflemire2017a,tofflemire2017b}. While these scenarios are not identical, a young star plus (potentially eccentric) hot Jupiter just inside the disk truncation radius is consistent with observations of CI~Tau, and these and similar simulations and observations suggest that the accretion onto the central star can show modulation of the flux or spectral features with the binary orbital period.  We suggest this is the source of the $\sim9.0$ d periodicity seen in the K2 data for CI~Tau.

\section{Summary}

Our periodogram analysis of the \ktwo photometry of CI~Tau reveals the presence of two persistent periodic signals at $\sim$6.6 d and $\sim$9.0 d.  We find that the $\sim$6.6 d signal likely corresponds to the stellar rotation period because in conjunction with our \vsini measurement, this value yields a reasonable estimate for the stellar radius, consistent with the effective temperature of a K7 star with CI~Tau's luminosity \citep{mcclure2013}.  If the rotation period is 9.0 d, the radius becomes unphysically large. The $\sim$9.0 d signal is consistent with the orbital period of a non-transiting, several Jupiter-mass planet located near the inner edge of the system's accretion disk \citep{johnskrull2016}.  We postulate that the $\sim$9.0 d signal originates from planet-disk interactions and the impact on the accretion of disk material onto the star is modulated by the planet's orbit and manifests in the system's photometric variability. The scenario by which a Jupiter-sized planet could influence accretion from a disk onto a star has not been simulated as far as we know. However, simulations of black hole binaries \citep{dorazio2013} have shown that with a mass ratio comparable to CI~Tau A and b, it is possible that visible accretion streams can be regulated with a period equal to the orbit of the binary.  Although this scenario does not simulate star-disk-planet interactions, it does suggest the possibility that a planet located within a disk's truncation radius can modulate accretion onto its host star with a period similar to its orbit. A model that simulates such a scenario, as is demonstrated in the CI~Tau system, would supplement our understanding of the physical processes that occur as planets interact with disk accretion streams onto young stars.

\section*{Acknowledgments}
We are grateful to S. Aigran, H. Parviainen, A. Vanderburg, and R. Luger for their valuable insight and assistance with \ktwo detrending procedures. We thank Wei Chen for their contribution to the \vsini we report in the paper. CMJ-K would like to acknowledge useful discussions with P. Hartigan. LIB wishes to thank L. Nofi for thoughtful discussion and insight. We would also like to extend our gratitude to the anonymous referee who provided valuable input that improved this work. This paper includes data collected by the \ktwo mission. Funding for the Kepler mission is provided by the NASA Science Mission directorate. All of the data presented in this paper were obtained from the Mikulski Archive for Space Telescopes (MAST). STScI is operated by the Association of Universities for Research in Astronomy, Inc., under NASA contract NAS5-26555. Support for MAST for non-HST data is provided by the NASA Office of Space Science via grant NNX09AF08G and by other grants and contracts.  This research has made use of NASA's Astrophysics Data System Bibliographic Services and the SIMBAD database, operated at CDS, Strasbourg, France.




\end{document}